\documentclass[twocolumn,prb,amsmath,amssymb,floatfix,superscriptaddress]{revtex4-2} 
\usepackage{amsmath}
\usepackage{amssymb}
\usepackage{graphicx}
\usepackage{bm}
\usepackage{color}
\usepackage{hyperref}
\usepackage{graphicx}
\usepackage{upgreek}
\usepackage{scalerel}
\usepackage{multirow}
 
\usepackage{pgffor}
\usepackage{pdfpages}
\usepackage{mathdots}
\usepackage{float}
\usepackage{silence}
\usepackage{physics} 
\usepackage{lscape}   
\usepackage{color, colortbl}
\usepackage[table]{xcolor}
\usepackage{comment}  
\makeatletter
\AtBeginDocument{\let\LS@rot\@undefined}
\makeatother

\usepackage{sidecap,tikz}
\definecolor{lime}{HTML}{A6CE39}
\DeclareRobustCommand{\orcidicon}{\hspace{-1mm}
	\begin{tikzpicture}
		\draw[lime, fill=lime] (0,0) 
		circle [radius=0.16] 
		node[white] {{\fontfamily{qag}\selectfont \tiny \,ID}};
		\draw[white, fill=white] (-0.0525,0.095) 
		circle [radius=0.007];
	\end{tikzpicture}
	\hspace{-3mm}
}
\foreach \x in {A, ..., Z}{\expandafter\xdef\csname orcid\x\endcsname{\noexpand\href{https://orcid.org/\csname orcidauthor\x\endcsname}
		{\noexpand\orcidicon}}
}

\begin{document}

\title{Non-Hermitian minimal Kitaev chains}
\author{Jorge Cayao\orcidA{}}
\email[]{jorge.cayao@physics.uu.se}
\affiliation{Department of Physics and Astronomy, Uppsala University, Box 516, S-751 20 Uppsala, Sweden}

\author{Ram\'{o}n Aguado\orcidB{}}
\email{ramon.aguado@csic.es}
\affiliation{Instituto de Ciencia de Materiales de Madrid (ICMM), Consejo Superior de Investigaciones Cient\'{i}ficas (CSIC), Sor Juana In\'{e}s de la Cruz 3, 28049 Madrid, Spain}

\date{\today}
\begin{abstract}
Starting from a double quantum dot realization of a minimal Kitaev chain, we  demonstrate that non-Hermiticity stabilizes the so-called poor man's Majorana zero modes in a region of parameter space that is much broader than in the Hermitian regime.  In particular, we consider the simplest non-Hermitian mechanism which  naturally appears due to coupling to normal reservoirs and is commonly present in all   transport experiments. Specifically, such couplings induce exceptional points which connect stable and highly tunable zero energy real lines that are well separated from the quasicontinuum.   Such zero-energy lines reflect spectral degeneracies protected by topology and represent the non-Hermitian generalization of the Hermitian poor mans Majorana modes occurring at single points in parameter space. Our findings pave the way for realizing robust non-Hermitian effects by combining unconventional superconductors and non-Hermitian topology.
 
\end{abstract}
\maketitle
\section{Introduction}
Majorana zero modes (MZMs) have attracted  enormous interest during the past decade not only because they characterize topological superconductivity \cite{tanaka2011symmetry,sato2016majorana,sato2017topological,cayao2019odd,tanakaReview2024}, a new state of matter, but also due to their potential use in fault tolerant quantum computation \cite{Sarma_NPJQ2015,10.1063/PT.3.4499,beenakker2019search}. This topological state was predicted to appear in enginereed $p$-wave superconductors, based on e.g. semiconductor-superconductor hybrid nanostructures \cite{Oreg_PRL2010,Lutchyn_PRL2010,AguadoReview,LutchynReview}. 

The unambiguous detection of MZMs has been hotly debated in the last decade \cite{Prada_review}. However,  recent experimental demonstration \cite{dvir2023realization,Haaf2024} of bottom-up engineering of a Kitaev chain \cite{Kitaev_2001} has given new impetus to the field.  These experiments confirm that a minimal Kitaev chain can be realized with only two quantum dots (QDs) coupled by a  superconductor \cite{Leijnse_PRB2012}; see also Ref.\,\cite{Sau_2012}. MZMs in these minimal chains only appear in fine-tuned ``sweet spots" in parameter space and without topological protection, so they are often called poor man's Majorana modes (PMMMs).  Reaching such sweet spots is challenging as it involves having equal  electron co-tunneling (ECT) and crossed Andreev reflection (CAR) \cite{PhysRevX.13.031031}, a condition that has  recently been achieved using a hybrid semiconductor-superconductor segment mediating the coupling between QDs \cite{dvir2023realization,Haaf2024};\cite{zatelli2023robust,Wang_Nat2022,Bordoloi_Nat2022,Wang_Natcom2023}.

Typical experiments    for detecting PMMMs involve local and nonlocal conductances \cite{dvir2023realization,Haaf2024}. As such, the sample is contacted by metallic leads which play the role of reservoirs, see e.g., Ref.\,\cite{dvir2023realization}. A suitable mathematical language to describe this open quantum system is the so-called non-Hermitian (NH) formulation \cite{Moiseyev,doi:10.1080/00018732.2021.1876991}. At very weak couplings, the NH impact of the leads slightly affects the sample via a broadening of the conductance measurements \cite{datta1997electronic}. In this limit, such measurements essentially reveal the spectrum of the closed quantum system. However, when the coupling to the leads is comparable to   other energy scales, NH effects  considerably change the properties of the isolated system and give rise to unique phenomena that does not exist in the Hermitian regime \cite{doi:10.1080/00018732.2021.1876991,RevModPhys.93.015005,OS23}; see also Refs.\,\cite{cayao2023non,li2024anomalous,shen2024nonhermitian,beenakker2024josephson,pino2024thermodynamics,cayao2024non}.

 \begin{figure}[!t]
\centering
	\includegraphics[width=\columnwidth]{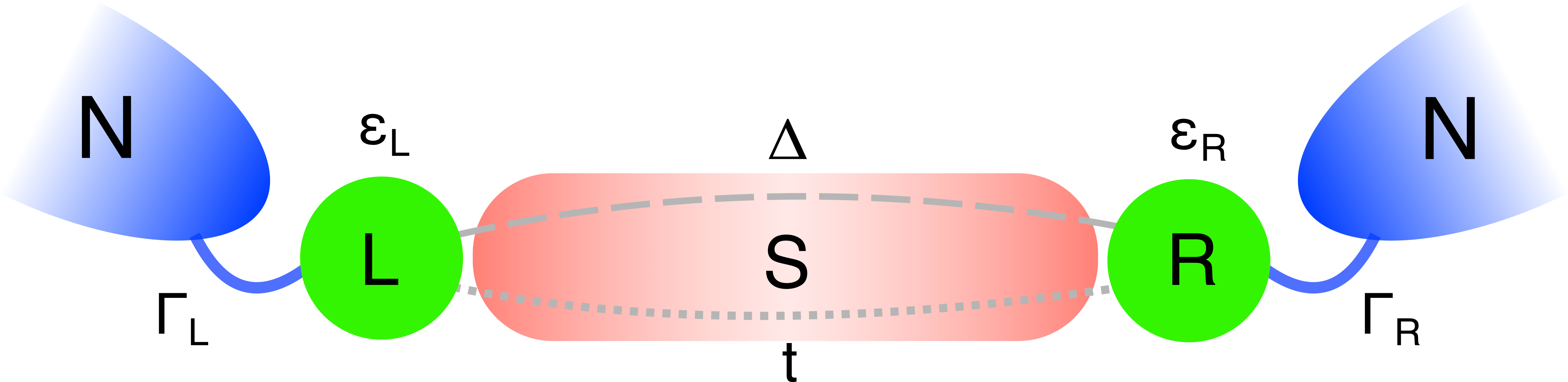}
 	\caption{A two-site  Kitaev chain  realized by coupling two QDs (green circles) with onsite energies $\varepsilon_{\rm L,R}$    through a common superconductor (red). This system hosts poor man's Majorana modes when single ($t$) and two electron ($\Delta$) tunneling are equal.
  Coupling this minimal Kitaev chain to normal leads N (blue regions) by rates $\Gamma_{\rm L,R}$ makes the system non-Hermitian.}
\label{Fig0} 
\end{figure}

In this paper, we demonstrate that non-Hermiticity is a powerful mechanism to stabilize and realize PMMMs through the emergence of NH spectral degeneracies  known as exceptional points (EPs), where eigenvalues and eigenvectors  coalesce. In particular,  we consider a NH two-site Kitaev chain  with non-Hermiticity due to coupling the two QDs to normal leads by $\Gamma_{\rm L,R}$; see Fig.\,\ref{Fig0}. We discover that, when  $\Gamma_{\rm L}\neq\Gamma_{\rm R}$, EP bifurcations generate zero-energy lines where PMMMs are stable beyond the Hermitian sweet spot. We further demonstrate that the emergent EPs and zero-energy lines can be detected via spectroscopy and conductance. Since EPs posses an inherent NH topology and are protected by fundamental symmetries \cite{PhysRevX.8.031079, PhysRevX.9.041015,RevModPhys.93.015005,doi:10.7566/JPSCP.30.011098,OS23}, the zero-energy lines and NH PMMMs represent topologically protected NH effects.

\section{NH two-site Kitaev chain}
We consider a minimal Kitaev model where two (left/right) QDs with energies $\varepsilon_{\rm L/R}$ couple via a superconductor that allows for CAR and ECT \cite{Leijnse_PRB2012,dvir2023realization,PhysRevX.13.031031}, see Fig.\,\ref{Fig0}. These two processes are parametrized by $\Delta$ and $t$, respectively, resulting in a minimal Kitaev Hamiltonian
 \begin{equation}
 \label{HK}
 H_{\rm K}=\varepsilon_{\rm L}\sigma_{+}\tau_{z}+\varepsilon_{\rm R}\sigma_{-}\tau_{z}+t\sigma_{x}\tau_{z}-\Delta\sigma_{y}\tau_{y}\,,
 \end{equation}
where $\sigma_{\pm}=(\sigma_{0}\pm\sigma_{z})/2$, and $\sigma_{i}$ ($\tau_{i}$) is the $i$-th Pauli matrix in the QD (Nambu) subspace. Despite the simplicity of the minimal Kitaev chain given by Eq.\,(\ref{HK}), there already exists experimental evidence of its realization in superconductor-semiconductor hybrids, see Refs.\,\cite{dvir2023realization,bordin2023crossed,Haaf2024,zatelli2023robust}.

Non-Hermiticity is introduced by coupling each QD to normal (N) leads, as in Fig.\,\ref{Fig0}. The resulting open system is described by an effective NH Hamiltonian that reads
\begin{equation}
\label{Heff1}
H_{\rm eff}=H_{\rm K}+\Sigma^{r}(\omega=0)\,,
\end{equation}
where $\Sigma^{r}(\omega=0)$ is the zero-frequency retarded self-energy characterizing the coupling of each QD to the N leads. Since only the imaginary  part of the self-energy induces NH effects \cite{pikulin2012topological,Ioselevich_2013,JorgeEPs,avila2019non,PhysRevResearch.1.012003,PhysRevB.105.094502,PhysRevB.105.155418,PhysRevB.107.035408,PhysRevB.107.104515,cayao2023non,cayao2023exceptional,PhysRevB.108.L041403,PhysRevB.109.L161404,cayao2024nonhermitian}, we take the
so-called wide limit \cite{datta1997electronic} and parametrize the couplings as $\Gamma_{\rm L/R}=\pi|\tau|^{2}\rho_{\rm L/R}$, where $\tau$ is the hopping amplitude and $\rho_{\rm L/R}$ the surface density of states of the left/right lead. This results in a purely imaginary self-energy of the form
$\Sigma^{r}(\omega=0)={\rm diag}(\Sigma_{e}^{r},\Sigma_{h}^{r})$, with $\Sigma_{e(h)}^{r}=-i\Gamma \sigma_{0}-i\gamma\sigma_{z}$, $\Gamma=(\Gamma_{\rm L}+\Gamma_{\rm R})/2$ and $\gamma=(\Gamma_{\rm L}-\Gamma_{\rm R})/2$.  This effective Hamiltonian approach  has been widely used to study quantum transport and it hence represents a realistic condensed matter ground to explore NH effects\cite{datta1997electronic}.  Interestingly, the effective Hamiltonian in Eq.\,(\ref{Heff1}) has the   particle-hole symmetry given by  $CH_{\rm eff}^{*}C^{\dagger}=-H_{\rm eff}$, also known as PHS$^{\dagger}$ symmetry,    where $CC^{*}=1$ and $C=\tau_{x}\sigma_{0}$ \cite{PhysRevLett.123.066405,cayao2023non}. Next we   explore the role of the PHS$^{\dagger}$ symmetry and  the NH self-energy for realizing PMMMs.

\section{PMMMs at the sweet spot $\Delta=t$}
We begin by analyzing the impact of non-Hermiticity on the sweet spot $\Delta=t$, which is  a single point in parameter space where PMMMs appear in the Hermitian regime ~\cite{dvir2023realization}. To inspect the realization of PMMMs, we obtain the eigenvalues of $H_{\rm eff}$. At $\varepsilon_{\rm L,R}=\varepsilon$, they are given by
\begin{equation}
\label{Eval1}
E_{j}^{\pm}=-i\Gamma\pm \sqrt{P-(-1)^{j}2\sqrt{Q}}\,,
\end{equation}  
where $P=2t^{2}+\varepsilon^{2}-\gamma^{2}$, $Q=t^{2}(t^{2}+\varepsilon^{2})-\varepsilon^{2}\gamma^{2}$, and $j=0(1)$ labels the   first (second) electron-  and hole-like eigenvalues. At  $\varepsilon=0$, we get $E_{0}^{\pm}=-i\Gamma\pm i|\gamma|$ and $E_{1}^{\pm}=-i\Gamma\pm \sqrt{4t^{2}-\gamma^{2}}$. 
Physically, the first set of eigenvalues corresponds to  PMMMs  which acquire different lifetimes, while the second set are bulk modes. \begin{figure}[!t]
\centering
	\includegraphics[width=0.5\textwidth]{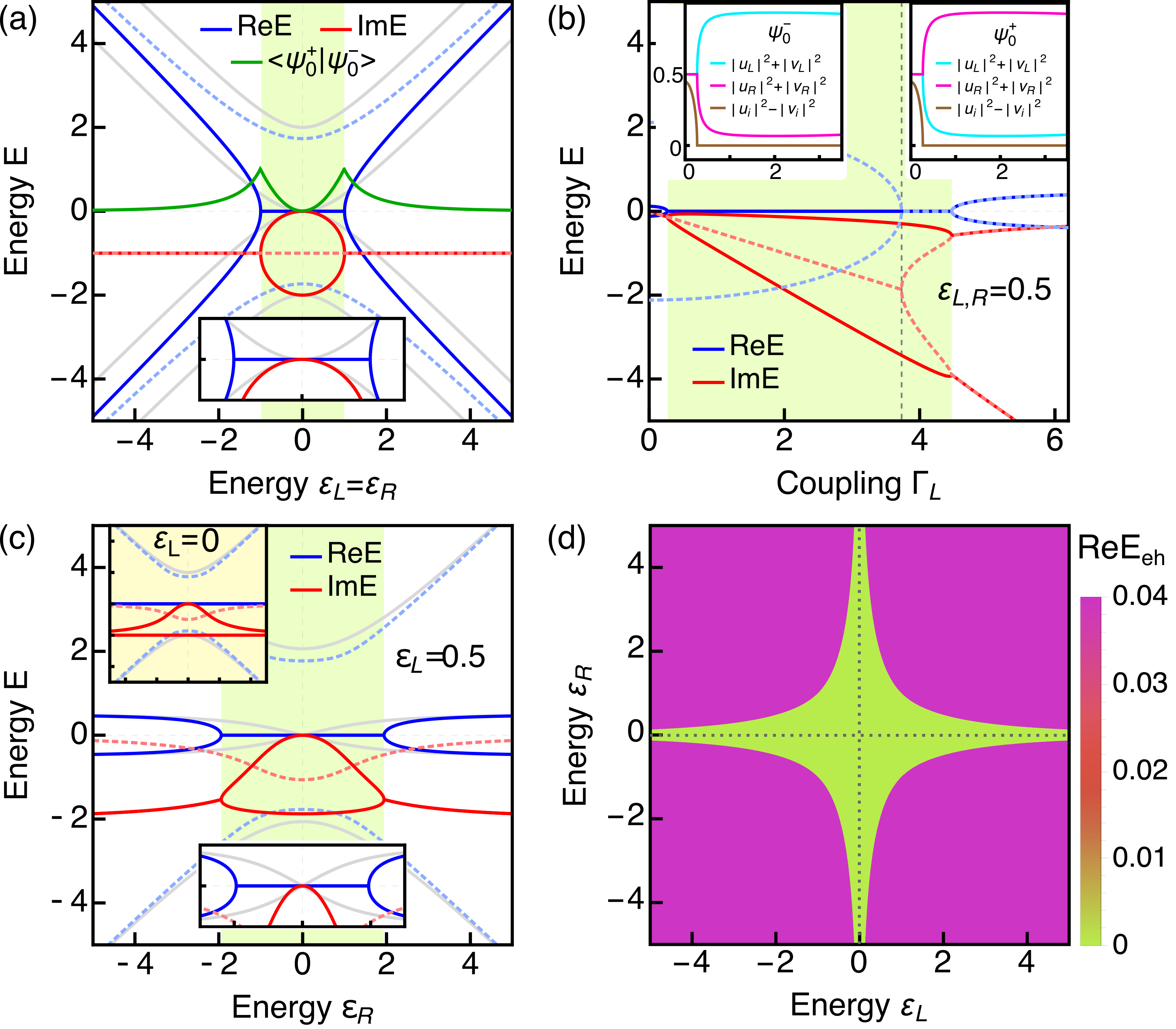}
 	\caption{Re (blue) and Im (red) parts of the eigenvalues in the sweet spot $\Delta=t$ at finite non-Hermiticity, where dashed light colors correspond to bulk eigenvalues. (a) The eigenvalues as a function of $\varepsilon_{\rm L}=\varepsilon_{\rm R}$, while (b,c) as a function of $\Gamma_{\rm L}$ at  $\varepsilon_{\rm L,R}=0.5$ and as a function of $\varepsilon_{\rm R}$ at $\varepsilon_{\rm L}=0.5$, respectively. The gray curves in (a-c)
  correspond to $\Gamma_{\rm L,R}=0$, while the green region ends   mark the EPs due to lowest eigenvalues.   The green curve in (a) shows the absolute value of the inner product between the wavefunctions of the lowest positive and negative states $\psi_{0}^{\pm}$. The cyan and magenta curves in the insets of (b) show the probability amplitudes of  $\psi_{0}^{\pm}$ in the left and right QDs as a function of $\Gamma_{\rm L}$, respectively, while the brown curve represents the quasiparticle charge associated to $\psi_{0}^{\pm}$. Here, $u_{i} (v_{i})$ represent the electron (hole) wavefunction components  of the $i={\rm L,R}$ QDs. Vertical dashed line in (c) marks   EPs due to higher energy eigenvalues. Panel (d) shows the Re part of the difference between lowest positive and negative eigenvalues ${\rm Re} E_{eh}$, where  the green region indicates ${\rm Re} E_{\rm eh}=0$ and its borders mark the EPs. The dotted lines indicate in (d) ${\rm Re} E_{\rm eh}=0$ at $\Gamma_{\rm L,R}=0$.   Parameters: $\Delta=t=1$, $\Gamma_{\rm L}=2$, $\Gamma_{\rm R}=0$.}
\label{Fig1} 
\end{figure}This implies that $E_{0}^{\pm}$ has zero real (Re) part and distinct imaginary (Im) components. We have verified that   the wavefunctions associated to these levels with zero Re part are well localized at each QD, in a similar fashion as in the Hermitian regime \cite{Leijnse_PRB2012}. Therefore, when  $\Delta=t$ and $\varepsilon=0$, non-Hermiticity only gives finite lifetimes to the already  localized zero-energy PMMMs.  In contrast, the bulk levels  $E_{1}^{\pm}$, have finite Re and Im parts, implying that there is always a gap for $2t>|\gamma|$. However, such gap vanishes at $2t=|\gamma|$ when $E_{1}^{\pm}$ merge into a single value. We have verified that the eigenvectors associated to  $E_{1}^{\pm}$ also merge, thus signaling the formation of  EPs at $2t=\pm |\gamma|$. Even though such EPs due to $E_{1}^{\pm}$ are inherently interesting, they are detrimental for  PMMMs, since there is no gap separating the zero modes from the quasicontinuum in real samples.

The situation becomes more interesting when  $\Delta=t$ but $\varepsilon\neq0$. 
In the Hermitian regime, the lowest energy levels develop a single zero-energy crossing at $\varepsilon=0$ which disperses as $E_{0}^{\pm}\approx\pm \varepsilon^2/(2\Delta)$, see gray curves in Fig.\,\ref{Fig1}(a), reflecting the so-called quadratic protection~\cite{Leijnse_PRB2012}. Interestingly, in the NH case, the lowest energy levels $E_{0}^{\pm}$, blue curves in Fig.\,\ref{Fig1}(a),  can acquire zero Re part for a large range of $\varepsilon\neq 0$ when $0<P\leq {\rm Re}(2\sqrt{Q})$ and ${\rm Im}(2\sqrt{Q})=0$, where the equality defines the formation of EPs and the inequality ensures zero Re part between EPs. From the equality, we obtain two QD energies at which EPs appear, $\pm \varepsilon_{\rm EP}^{\pm}=\pm\sqrt{\gamma(\pm2t-\gamma)}$, giving only two EPs at $\pm \varepsilon_{\rm EP}^{+(-)}$ for $\gamma>0$ ($\gamma<0$). These EPs connect lines with zero Re energy formed by $E_{0}^{\pm}$, while their Im parts split in a circular fashion between EPs, see Fig.\,\ref{Fig1}(a). This is a remarkable effect of non-Hermiticity on the lowest states, which clearly enforces them to develop a line of zero energy points in stark contrast to the Hermitian regime, where the lowest levels acquire zero energy only at a single point.  Unlike the case with $\varepsilon=0$,  here  $E_{1}^{\pm}$ at $\pm \varepsilon_{\rm EP}^{+}$ remain  finite with constant Im parts and energy-dependent Re  parts which  result in a finite gap between the lowest energies and the bulk; see Fig.\,\ref{Fig1}(a).   As expected, the wavefunctions $\psi_{0}^{\pm}$ associated to $E_{0}^{\pm}$ become parallel at the EPs, as reflected by their inner product reaching unity,  $\langle \psi_{0}^{+}|\psi_{0}^{-}\rangle=1$; see green curve in  Fig.\,\ref{Fig1}(a). It is also evident from $\varepsilon_{\rm EP}^{+}$ that larger values of $t$  make the regions with zero Re part longer. However,  while  a finite $\gamma$ is necessary to induce such regions,  $\gamma>2t$ is detrimental because $\varepsilon_{\rm EP}^{+}$ becomes Im. In this case, large  $\gamma$ affects the bulk levels $E_{1}^{\pm}$, making them   develop EPs as well; see Fig.\,\ref{Fig1}(b).   It is also worth noting that, while  $\varepsilon\neq0$ gives delocalized wave functions in the Hermitian regime \cite{Leijnse_PRB2012}, after the EP transition,  $\psi_{0}^{\pm}$ become strongly localized in the right (left) QD and develop vanishing quasiparticle charge characterized by $|u_{i}|^{2}-|v_{i}|^{2}$; see magenta, cyan, and brown curves in the insets of Fig.\,\ref{Fig1}(b). This behavior   supports the Majorana charge neutrality nature of the zero Re energy states found here.  We stress that the EPs and their nontrivial NH effects discussed here require  asymmetric couplings to the normal leads; this is because symmetric couplings only produce eigenvalues with the same Im part which do not form EPs.

For $\varepsilon_{\rm L}\neq\varepsilon_{\rm R}$, the bowtie dependence \cite{PhysRevB.96.085418,PhysRevResearch.5.043182} of the lowest levels with a single zero-energy crossing  in the Hermitian regime transforms into a line with zero Re part  and split Im parts with a bubble-like profile under non-Hermiticity [Fig.\,\ref{Fig1}(c)]. Further insight is obtained from Fig.\,\ref{Fig1}(d), where we plot a phase diagram expressed as the Re part of the difference between the lowest electron- and hole-like eigenvalues ${\rm Re}E_{\rm eh}={\rm Re}(E^{+}_{0}-E_{0}^{-})$. Here the green region indicates  ${\rm Re}E_{\rm eh}=0$, while its borders mark the EPs; the dotted horizontal and vertical lines show ${\rm Re}E_{\rm eh}=0$ in the Hermitian regime. A finite non-Hermiticity $\gamma\neq0$ induces large regions where the lowest two energy levels become zero modes. As a consequence, non-Hermiticity  at the sweet spot $\Delta=t$ offers a much broader range of parameter space to  realize PMMMs.

To end this section, we note that the emergence of EPs connecting the zero Re energy lines discussed above is a NH topological phenomenon protected by PHS$^{\dagger}$ symmetry \cite{PhysRevLett.123.066405,cayao2023non}. This implies that the   EPs found here are protected by a zero-dimensional $Z_{2}$ topological number, which corresponds to the  $D^{\dagger}$  NH topological class  \cite{PhysRevLett.123.066405}. The nontrivial NH topological aspect of the formation of EPs can be also understood by noting that  PHS$^{\dagger}$ symmetry here implies that if $E_{i}$ is an eigenvalue of the effective NH Hamiltonian, then also $-E_{i}^{*}$ is an eigenvalue.  Interestingly, this symmetry between eigenvalues can be satisfied in two different manners: (i) The PHS$^{\dagger}$ symmetry can be satisfied for pairs of eigenvalues with the same imaginary part $(E_{+},E_{-})$, where $E_{\pm}=\pm E-i\Gamma$,   with $E_{\pm}=-E_{\mp}^{*}$, and the label $\pm$ denotes the  electron and hole energy levels. This situation  corresponds to the positive and negative eigenvalues closest to zero Re energies before getting degenerated in, e.g., Fig.\,\ref{Fig1}(a); (ii) the PHS$^{\dagger}$ symmetry can be also satisfied for nondegenerate self-conjugate imaginary eigenvalues   $E_{\pm}=-i\Gamma_{\pm}$, which implies  $E_{\pm}=-E_{\pm}^{*}$.  We thus see that, within this process,  two energy levels  following $E_{\pm}=-E_{\mp}^{*}$ bifurcate into two modes having zero Re energy and distinct lifetimes following $E_{\pm}=-E_{\pm}^{*}$, signaling a nontrivial phenomenon that defines the emergence of an EP in NH systems. This therefore implies that a non-degenerate energy   on the imaginary axis cannot acquire a nonzero Re part without breaking the self-conjugation symmetry due to PHS$^{\dagger}$  \cite{pikulin2012topological,PhysRevB.87.235421,Ioselevich_2013,RevModPhys.87.1037,JorgeEPs,avila2019non}. We stress that the self-conjugation symmetry imposed by PHS$^{\dagger}$ is preserved along the zero Re energy lines connecting EPs, unveiling their intrinsic NH topological origin.  It is important to remark that, due to the inherent swapping of eigenvalues and eigenstates at EPs \cite{RevModPhys.93.015005}, NH PMMMs along the zero Re energy   lines are very likely to exhibit nontrivial braiding-like properties entirely protected by NH topology.

\begin{figure}[!t]
\centering
	\includegraphics[width=0.5\textwidth]{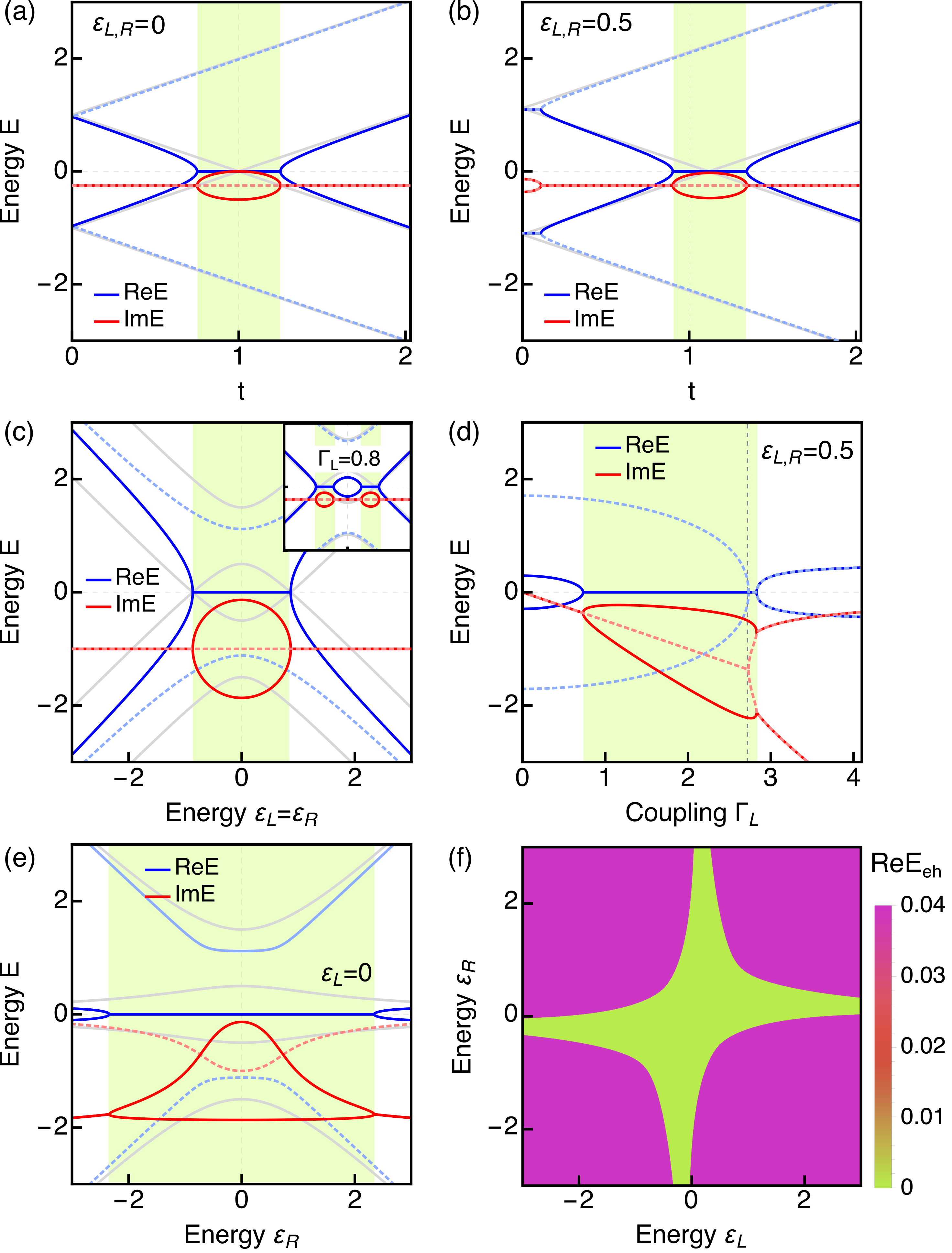}\\
 	\caption{Re (blue) and Im (red) parts of the eigenvalues 
    away from the sweet spot $\Delta\neq t$ at finite non-Hermiticity.  (a,b) Eigenvalues as a function of $t$ for $\varepsilon_{\rm L,R}=0$ and $\varepsilon_{\rm l,R}=0.5$ at $\Delta=1$ and $\Gamma_{\rm L}=0.5$ and $\Gamma_{\rm R}=0$.    (c) Eigenvalues as a function of $\varepsilon_{\rm L}=\varepsilon_{\rm R}$ at $\Gamma_{\rm L}=2$ and $\Gamma_{\rm R}=0$, with the inset at $\Gamma_{\rm L}=0.8$.  (d) Eigenvalues as a function of $\Gamma_{\rm L}$ at  $\varepsilon_{\rm L,R}=0.5$. 
      (e) depicts the eigenvalues as a function of   $\varepsilon_{\rm R}$ at $\varepsilon_{\rm L}=0$  at $\Gamma_{\rm L}=2$ and $\Gamma_{\rm R}=0$.   The gray curves in (a-e) correspond to $\Gamma_{\rm L,R}=0$, while the green region ends mark the EPs due to lowest eigenvalues. Vertical dashed lines in (e) mark   EPs due to higher energy eigenvalues.   (f)   Re part of the difference between lowest positive and negative eigenvalues ${\rm Re} E_{eh}$, where  the green region indicates ${\rm Re} E_{\rm eh}=0$ and its borders mark the EPs.   Parameters: in (c-f) $\Delta=0.5$, $t=1$, $\Gamma_{\rm L}=2$, $\Gamma_{\rm R}=0$.}
\label{Fig2} 
\end{figure}

\section{Away from the sweet spot $\Delta\neq {t}$}
Relaxing the ideal sweet spot conditions to $\Delta\neq t$, the eigenvalues at $\varepsilon_{\rm L,R}=\varepsilon$ are still given by Eqs.\,(\ref{Eval1})   but now with $P=t^{2}+\Delta^{2}+\varepsilon^{2}-\gamma^{2}$ and $Q=t^{2}(\Delta^{2}-\varepsilon^{2})-\varepsilon^{2}\gamma^{2}$.  In this case, the eigenvalues develop EPs  when $0<P\leq {\rm Re}(2\sqrt{Q})$ and ${\rm Im}(2\sqrt{Q})=0$, with the equality defining the EP positions as a function of $t$ given by $ \pm t^{\pm}_{\rm EP}=\pm\sqrt{(\Delta\pm \gamma)^{2}+\varepsilon^{2}}$; or as a function of the onsite energy $\pm \varepsilon _{\rm EP}^{\pm}=\pm \sqrt{t^{2}-(\Delta\pm\gamma)^{2}}$.    Unlike the sweet spot case, here there can be four EPs at $\pm t^{\pm}_{\rm EP}$ or $\pm \varepsilon _{\rm EP}^{\pm}$, two at negative and two at positive $t$ or $\varepsilon$, provided $t>|\Delta\pm \gamma|$ for the latter case. Even for small NH asymmetries $\gamma$, the single zero-energy Hermitian crossing at $t=\Delta$  splits into two EPs $t^{\pm}_{\rm EP}$, resulting in zero modes for $t\neq \Delta$, an effect that occurs for both vanishing and finite $\varepsilon$, see Figs.\,\ref{Fig2}(a,b).  As noted, the  EPs also appear as a function of $\varepsilon$, where four EPs form at small $\gamma$ but, as it gets larger, the EPs closer to $\varepsilon=0$ fuse, leaving only the outer EPs connecting lines with zero Re part and circular Im parts ${\rm Im}E_{0}^{\pm}\neq0$, see Figs.\,\ref{Fig2}(c) and inset therein.   The stabilization of $E_{0}^{\pm}$ at zero Re energy by a finite non-Hermiticity is very similar to what we obtain in the sweet spot, but, of course, here no zero energy PMMMs are expected in the Hermitian regime. We note that the zero-energy line between EPs is well separated from the higher energy levels $E_{1}^{\pm}$, which provides an excitation gap. However,  very large non-Hermiticity is not useful because it can force  $E_{1}^{\pm}$ to zero Re energy via an EP transition; see Fig.\,\ref{Fig2}(d).

When $\varepsilon_{\rm L}\neq\varepsilon_{\rm R}$, the formation of zero energy Re lines between  EPs in  $E_{0}^{\pm}$ remains robust; see  Fig.\,\ref{Fig2}(e).
At $\gamma=0$, the eigenvalues develop a diamond-like profile which  transforms into a line with zero Re energy whose ends mark the EPs, with a bubble-like splitting of the Im parts similar to the sweet spot in Fig.\,\ref{Fig1}(c). The appearance of zero-energy Re lines is a stable effect occurring in a large set of parameters;  see Fig.\,\ref{Fig2}(f);  the green region indicates ${\rm Re}E_{\rm eh}=0$ and its borders signal the position of EPs.  We thus conclude that non-Hermiticity has the potential to  induce and stabilize zero-energy  PMMMs in parameter regimes well beyond the Hermitian case.

\section{Many-body description under non-Hermiticity}
\begin{figure}[!t]
\centering
	\includegraphics[width=0.5\textwidth]{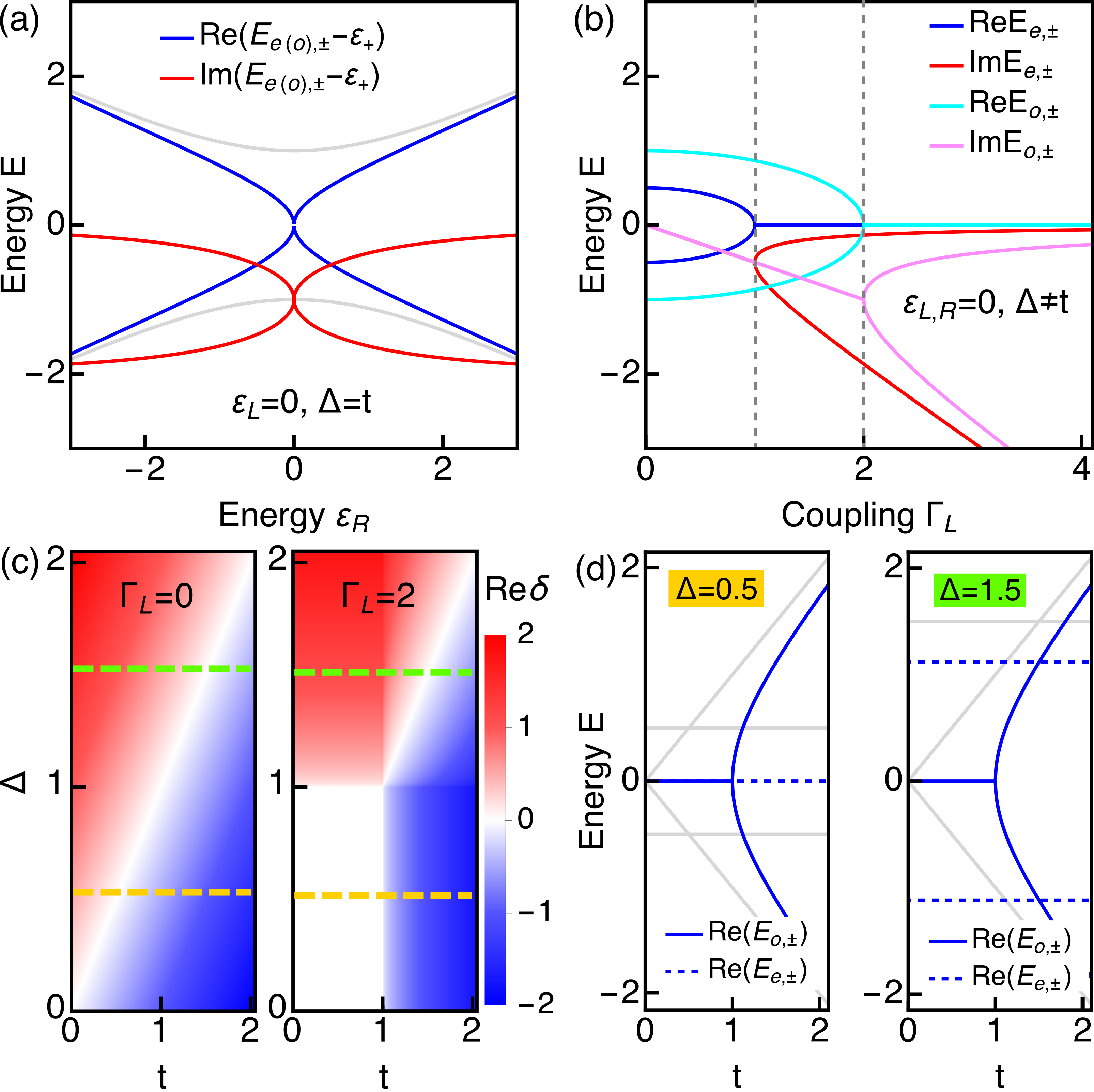}\\
 	\caption{(a) Re   and Im  parts of the many-body eigenvalues at finite non-Hermiticity as a function of $\varepsilon_{\rm R}$ at $\Delta=t=1$, while (b) as a function of $\Gamma_{\rm L}$ at $\Delta=0.5$ and $t=1$. Gray curves in (a) are the Hermitian energies; vertical dashed lines in (b) mark the EPs. (c) ${\rm Re}(\delta)=Re(E_{o,-}-E_{e,-})$ as a function of $\Delta$ and $t$ at  $\Gamma_{\rm L/R}=0$ and $\Gamma_{\rm L}=2$, $\Gamma_{\rm R}=0$; here $\varepsilon_{\rm L,R}=0$. White region indicates ${\rm Re}\delta=0$. (d) Re parts of the many-body eigenvalues at two fixed values of $\Delta$ in (c), indicated by orange and green dashed lines.   Parameters:   $\Gamma_{\rm R}=0$.}
\label{Fig3} 
\end{figure}

Further understanding of the role of non-Hermiticity is obtained by inspecting the many-body energies. We thus write $H_{\rm eff}$ given by Eq.\,(\ref{Heff1}) in the occupation number basis $\ket{n_{L}n_{R}}$, with $n_{L,R}$ being the electron number on each QD. Hence, $\mathcal{H}_{\rm eff}={\rm diag}(H_{e},H_{o})$ for the even (e) and odd (o) occupations {($\ket{00},\ket{11},\ket{01},\ket{10}$)}. Here,   $H_{e}=(\varepsilon_{+}-i\Gamma)(\eta_{0}-\eta_{z})+\Delta \eta_{x}$ and $H_{o}=(\varepsilon_{+}-i\Gamma)\bar{\eta}_{0}+(\varepsilon_{-}-i\gamma)\bar{\eta}_{z}+t \bar{\eta}_{z}$, where $\varepsilon_{\pm}=(\varepsilon_{\rm L}\pm\varepsilon_{\rm R})/2$, and $\eta_{i}$ ($\bar{\eta}_{i}$) is the $i$-th Pauli matrix in the even (odd) occupation basis subspace. The eigenvalues of $H_{e(o)}$ are then given by
\begin{equation}
\label{Eval2}
\begin{split}
E_{e,\pm}&=-i\Gamma +\varepsilon_{+}\pm\sqrt{\Delta^{2}+(\varepsilon_{+}-i\Gamma)^{2}}\,,\\
E_{o,\pm}&=-i\Gamma +\varepsilon_{+}\pm\sqrt{t^{2}+(\varepsilon_{-}-i\gamma)^{2}}\,.
\end{split}
\end{equation}
As expected, Eqs.\,(\ref{Eval2}) reveal that the even sector depends on $\Delta$ (describing CAR), while the odd sector on  $t$ (describing ECT). A less obvious consequence of coupling this many-body basis to reservoirs is that EPs in the even and odd sectors are governed by $\Gamma$ and $\gamma$, respectively. Specifically, EPs in the even sector form when   $E_{e,\pm}$ merge into a single value, namely, when the square root vanishes, $\varepsilon_{+}=0$ and $\Delta^{2}=\Gamma^{2}$; even at equal couplings to the leads $\gamma=0$.  In contrast, for the odd sector, $E_{o,\pm}$ merge into a single value when $\varepsilon_{-}=0$ and $t^{2}=\gamma^{2}$, thus always requiring $\gamma\neq0$. At $\varepsilon_{\rm L,R}=0$, the EPs of both sectors give zero Re parts and finite Im energies. The EPs discussed here can be seen in Fig.\,\ref{Fig3}(a,b), where we   plot $E_{e(o),\pm}$   as a function of $\varepsilon_{\rm R}$ and $\Gamma_{\rm L}$. The even and odd sectors develop  EPs at  finite non-Hermiticity and they require different amounts of $\Gamma_{\rm L,R}$ away from $\Delta=t$. 

Having shown the emergence of EPs in the many-body description, a natural question  is what happens to the PMMMs discussed in the previous section, namely   to the zero-energy excitations between the degenerate even and odd ground states \cite{Leijnse_PRB2012}, which here can be characterized by ${\rm Re}\delta={\rm  Re}(E_{o,-}-E_{e,-})=0$. In the Hermitian regime, $\Gamma_{\rm L,R}=0$, the condition $\delta=0$ leads to the sweet spot $\varepsilon_{\rm L}\varepsilon_{\rm R}=t^{2}-\Delta^{2}$. Under non-Hermiticity,   ${\rm Re}\delta=0$ implies that  the sweet spot can be generalized as 
 \begin{equation}
 \label{NHSweetSpot}
 \begin{split}
 \varepsilon_{\rm L}\varepsilon_{\rm R}&=(t^{2}-\gamma^{2})-(\Delta^{2}-\Gamma^{2})\,,\\
 \Gamma\varepsilon_{+}&=\gamma\varepsilon_{-}\,.
 \end{split}
 \end{equation}
These conditions imply that, to achieve ${\rm Re}\delta=0$, it is necessary that both sectors undergo EP transitions. While this is not immediately obvious from Eqs.\,(\ref{NHSweetSpot}), a clearer expression for $\delta$ can be obtained for $\varepsilon_{\rm L,R}=0$, which gives 
\begin{equation}
\label{NHSweetdelta}
 \delta=-\sqrt{t^{2}-\gamma^{2}}+\sqrt{\Delta^{2}-\Gamma^{2}}\,.   
\end{equation}
The first (second) square root corresponds to the odd (even) sector  energy,   which, at $\Delta\neq t$, host   EP transitions  at different $\Gamma_{\rm L,R}$. Importantly, Eq.\,(\ref{NHSweetdelta}) reveals that ${\rm Re}\delta=0$, associated to the energy of PMMMs, can be only achieved if \emph{both} square roots become Im, which only happens after EP transitions in both sectors. To illustrate this effect, we plot in Fig.\,\ref{Fig3}(c) ${\rm Re}\delta$  as a function of $\Delta$ and $t$.  The white color indicates ${\rm Re}\delta=0$, which only occurs along a single line $\Delta=t$ in the Hermitian regime (left panel); this stems from  a single point parity crossing of the even and odd energies, see gray lines  in Fig.\,\ref{Fig3}(d). Notably, in the NH regime, ${\rm Re}\delta=0$ emerges in a wider region of the $\Delta-t$ plane [Fig.\,\ref{Fig3}(c), right panel], which here follows from Eq.\,(\ref{NHSweetdelta}) and reveals that the edges of such a white region are determined by $t\leq\gamma$ and $\Delta\leq\Gamma$, with the equality given the EP transitions and the inequality ensuring zero-energy PMMMs.


\section{Spectral and conductance signatures}
Having established the stabilization  and the realization of zero-energy states by EPs under non-Hermiticity, here we inspect their impact on the spectral function $A$ and  differential conductance $G_{\alpha\beta}=dI_{\alpha}/dV_{\beta}$. The former is obtained as $A(\omega)=-{\rm Im}Tr(\mathcal{G}^{r}-\mathcal{G}^{a})$, where $\mathcal{G}^{r}(\omega)=[\mathcal{G}^{a}(\omega)]^{\dagger}=(\omega-H_{\rm eff})^{-1}$ is the retarded  Green's function associated to  $H_{\rm eff}$  given by Eq.\,(\ref{Heff1}). The conductance is obtained as $G_{\rm \alpha\beta}(\omega)=(e^{2}/h)(\delta_{\alpha\beta}-|S_{ee}^{\alpha\beta}(\omega)|^{2}+|S_{he}^{\alpha\beta}(\omega)|^{2})$, where  $S_{ij}^{\alpha\beta}$ are elements of the $S$-matrix in Nambu space $S=1-iW^{\dagger}\mathcal{G}^{r}(\omega) W$, with  $W={\rm diag}\{\sqrt{\Gamma_{\rm L}},\sqrt{\Gamma_{\rm R}},-\sqrt{\Gamma_{\rm L}},-\sqrt{\Gamma_{\rm R}}\}$  characterizing the coupling to the L/R leads. 

\begin{figure}[!t]
\centering
	\includegraphics[width=0.48\textwidth]{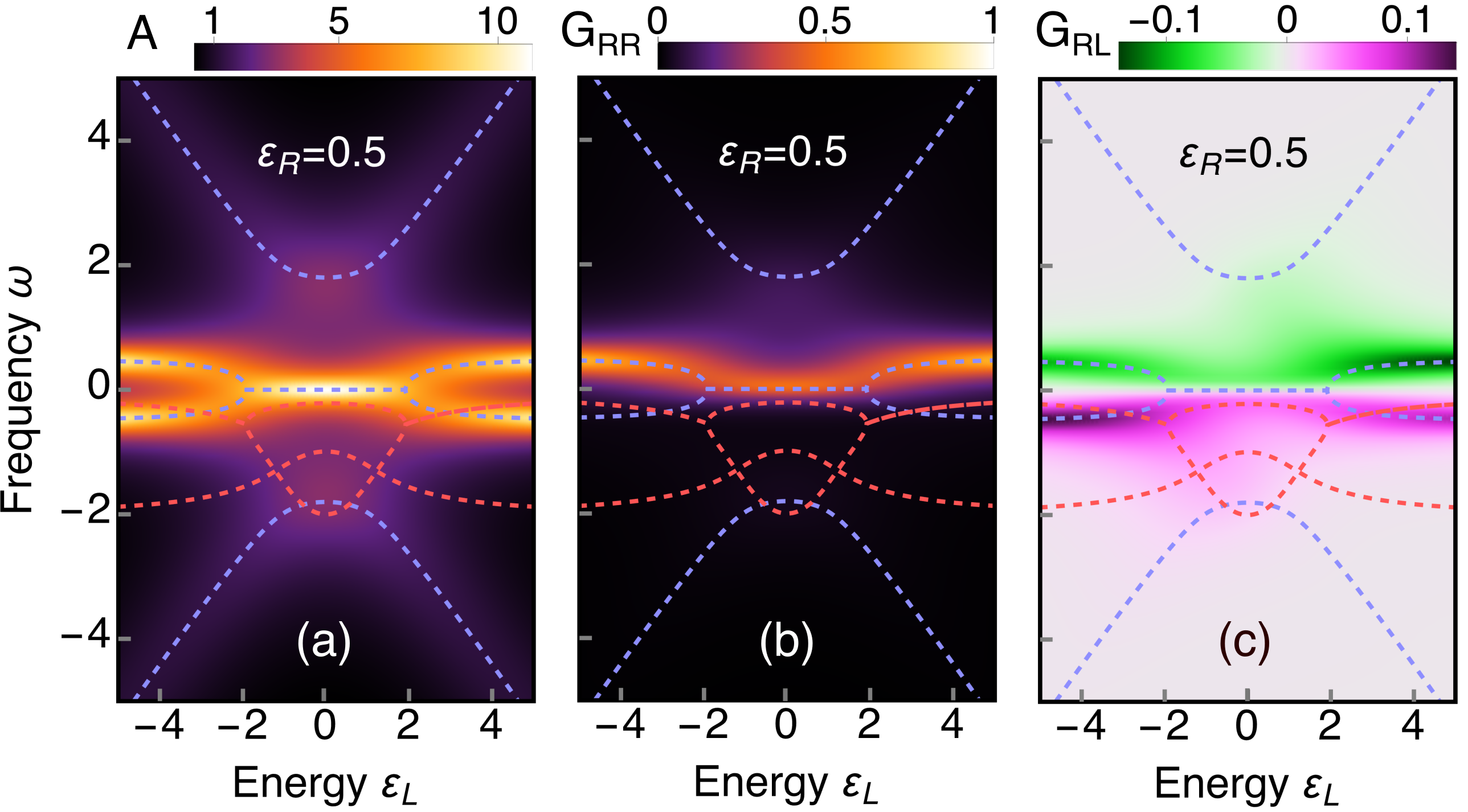}\\
 	\caption{(a) Spectral function, (b) local conductance, and (c) nonlocal conductance as functions of frequency  $\omega$ and onsite energy $\varepsilon_{\rm L}$ at $\Delta=t$ and $\Gamma_{\rm L}=2, \Gamma_{\rm R}=0.1$. Dashed blue and red curves in all panels represent the Re and Im parts of the eigenvalues. Parameters: $\Delta=1$, $t=1$, $\varepsilon_{\rm R}=0.5$.}
\label{Fig4} 
\end{figure}

In Fig.\,\ref{Fig4} we present $A$, $G_{\rm RR}$, and $G_{\rm RL}$ as a function of frequency $\omega$ and onsite energy $\varepsilon_{\rm L}$ at $\Delta=t$ for distinct couplings to the N leads and $\varepsilon=0.5$; $G_{\alpha\beta}$ is shown in units of $(e^{2}/h)$. The Re and Im parts of the eigenvalues are also shown in dashed blue and red curves. The main observation is that these quantities reveal the formation of EPs   and also the stable zero Re energy lines connecting them. The spectral function exhibits large values at low energies, giving rise to two clearly visible   Lorentzian resonances centered at positive and negative  frequencies, $\omega={\rm Re}E_{0}^{\pm}$, with their height and width determined by ${\rm Im}E_{0}^{\pm}$.  Notably, at the EPs, the two Lorentzian resonances merge into a single resonance centered at $\omega=0$, which then sticks at zero frequency and acquires large spectral weight between EPs  [Fig.\,\ref{Fig4}(a)]. The profile of the spectral function across EPs is therefore the key for detecting them and also for identifying the stable zero energy Re lines with NH PMMMs. Similar behavior is observed in the local conductance $G_{\rm RR}$, where the positive resonance gets to $\omega=0$ at the EPs but, in contrast to the case with symmetric $\Gamma_{\rm L/R}$, here $G_{\rm RR}$ is not necessarily quantized due to $\varepsilon_{\rm R}\neq0$, see  Fig.\,\ref{Fig4}(b). When it comes to the nonlocal conductance $G_{\rm RL}$, it exhibits positive and negative large values for $\omega<0$ and $\omega>0$, whose intensities reduce at the EPs, thus enabling the EP detection even though at $\omega=0$ the nonlocal conductance vanishes.

\section{Conclusions}
In conclusion, we have demonstrated that non-Hermiticity  represents a powerful mechanism for stabilizing   poor man's Majorana modes in  minimal Kitaev chains. In particular, we have considered a two-site Kitaev chain under non-Hermiticity due to coupling to normal leads and discovered that non-Hermiticity induces topologically protected exceptional points, connected by robust zero energy real lines that represent the non-Hermitian realization of the Hermitian poor man's Majorana modes.  Our results may help guide the  stabilization of   poor man's Majorana modes and pave the way for  realizing  stable non-Hermitian effects by combining few-site Kitaev chains and non-Hermitian topology. Specifically, our findings hold particular relevance for recent experiments, where few-site Kitaev chains have been successfully fabricated \cite{dvir2023realization,bordin2023crossed,Haaf2024,zatelli2023robust} and the precise control of couplings to normal leads seems to be feasible \cite{chen2023gate,PhysRevX.13.031031,Wang_Nat2022}.

\section{Acknowledgements}
We thank   M. Sato  and R. Seoane for insightful discussions.   J. C. acknowledges financial support from the Swedish Research Council (Vetenskapsr{\aa}det Grant No. 2021-04121),  the Royal Swedish Academy of Sciences (Grant No. PH2022-0003), and the Carl Trygger’s Foundation (Grant No. 22: 2093).  R. A. acknowledges support from the Horizon Europe Framework Program of the European Commission through the European Innovation Council Pathfinder Grant No. 101115315 (QuKiT), the Spanish Ministry of Science through Grants PID2021- 125343NB-I00 and TED2021-130292B-C43 funded by MCIN/AEI/10.13039/501100011033, ``ERDF A way of making Europe" and European Union NextGenerationEU/PRTR. Support by the CSIC Interdisciplinary Thematic Platform (PTI+) on Quantum Technologies (PTI-QTEP+) is also acknowledged.

\bibliography{biblio}

\end{document}